\documentstyle[prl,aps,multicol,amssymb,epsf]{revtex}
\begin{document}

\title{\boldmath Multiple-scattering effects on
        smooth neutron spectra}

\author{Joachim Wuttke}
\address{Physik-Department E13, Technische Universit\"at M\"unchen, 
         85747 Garching, Germany}

\date{submitted to Phys.~Rev.~E (23feb00), resubmitted (25apr00), 
      cond-mat/0002363}

\maketitle

\begin{abstract}
Elastic and inelastic incoherent neutron scattering experiments are simulated 
for simple models:
a rigid solid (as used for normalisation),
a glass (with a smooth distribution of harmonic vibrations),
and a viscous liquid (described by schematic mode-coupling equations).
As long as the spectral distribution
of the input scattering law does not vary with wavenumber,
it is only weakly distorted by multiple scattering.
The wavenumber dependence of the scattering intensity
suffers much more.
\end{abstract}

\begin{multicols}{2}

\section{Introduction}

Any neutron scattering measurement is 
unavoidably contaminated by multiple scattering.
For intensity reasons, samples must be chosen so thick
that a significant fraction of the incident neutrons is scattered.
As an inevitable consequence, a significant fraction of the scattered neutrons
is scattered more than once.

In crystals, single scattering gives rise to discrete peaks
that can be distinguished fairly well from
a smooth background caused by multiple scattering.
In amorphous solids and liquids, on the other hand,
the dynamic structure factor $S(q,\nu)$ itself is a smooth function
of wavenumber $q$ and frequency $\nu$.
In this case, 
the multiple-scattering background cannot be removed by routine operations,
and often it presents the limiting uncertainty in the data analysis.

Multiple scattering is basically a convolution of $S(q,\nu)$ with itself, 
and therefore it is nonlinear in~$S$, 
and worse: it is nonlocal in $q$ and $\nu$.
For this reason, multiple-scattering corrections are much more difficult
than all the other manipulations that are necessary
for deriving $S(q,\nu)$ 
from the counts $N(2\theta,\nu)$ measured at given detector angles $2\theta$:
normalisation to the incident flux,
subtraction of container scattering,
correction for self-absorption,
calibration to an incoherent standard scatterer,
correction for the energy-dependent detector efficiency,
and interpolation from constant-$2\theta$ to constant-$q$ cuts.

The nonlinearity of multiple scattering
means that any correction requires $S(q,\nu)$ to be known in absolute units.
The nonlocality means that
a multiple-scattering event registered in a channel $2\theta,\nu$
results from a succession of scattering events 
at other angles and frequencies $2\theta_i,\nu_i$ ($i=1,2,\ldots$).
Corrections are only possible if $S(q,\nu)$ 
is known over a wide range in $q$ and $\nu$.
Some of the multiple-scattering sequences that contribute to $N(2\theta,\nu)$
involve even angles or frequencies that are not covered directly
in the given experiment.
Therefore, it is impossible
to infer the distribution of multiple-scattering 
from the measured $N(2\theta,\nu)$ alone.
A full treatment of multiple scattering 
requires an extension of the measured scattering law 
into a wider $q,\nu$ domain.

In a pragmatic approach, 
this extension is provided
either by somehow extrapolating the measured data or 
by fitting a more or less physical model to them.
Feeding the extended scattering law into a simulation
one can estimate the multiple-scattering contribution,
and subtract it from the measured data.
After a few iterations one expects to obtain
a reasonably corrected scattering law.
Though such a procedure is regularly employed by a number of researchers,
it never became part of the standard raw data treatment.
The technical intricacies and inherent uncertainties 
of multiple-scattering corrections are rarely discussed in detail,
and for the uninitiated it is almost impossible to assess their reliability.

The present work follows an alternative route:
by performing extensive simulations on very simple model systems
we shall try to identify some generic trends of multiple scattering.
Ideally our results will help to assess past experiments 
 and to plan future ones.
Since we do not intend to correct data from a specific measurements,
we choose the simplest sample geometry,
and we do not consider scattering from the container.

We expect multiple scattering 
to be particularly harmful when 
the scattering law varies only weakly with $q$ and~$\nu$,
because small distortions of $S(q,\nu)$ suffice to destroy 
much of the information we are interested in.
To investigate such situations,
we consider incoherent scattering from a number of dynamic models.
The scattering laws will be defined by closed mathematical expressions
that cover the full $q,\nu$ plane, thereby guaranteeing correct normalisation.
To keep the models in touch with reality,
the choice of parameters will be inspired by actual experiments
 on organic glasses and liquids.

We start with simulating the vanadium or low-temperature scans
needed for normalisation of the elastic scattering intensity.
We then proceed with elastic and inelastic scattering
from a simple harmonic system.
This case has already been discussed more or less explicitely in 
experimental studies of amorphous solids
\cite{BuND84,CuDo90,WuKB93,BuPZ96a,SeDo97}.

In liquids, diffusion or slow relaxation cause the elastic part of
the scattering law to broaden into a quasielastic $\alpha$ peak.
Multiple-scattering effects in this regime have been studied occasionally 
\cite{Bee88,WuCR96b,Zor97b}.
More recently, interest has grown in the moderately viscous state
above the cross-over temperature $T_{\rm c}$ of mode-coupling theory
\cite{Got91,GoSj92}
where a relatively narrow $\alpha$ peak is separated from the
vibrational and relaxational high-frequency spectra
by the intermediate regime of fast $\beta$ relaxation.
By explicit integration of a schematic mode-coupling model
we construct an $S(q,\nu)$ 
which can be used as input to the multiple-scattering simulation.

\section{\boldmath Modelling $S(q,\nu)$}


\subsection{Rigid model} \label{MR}

The {\it rigid} model represents a completely frozen,
perfectly incoherent scatterer
\begin{equation}\label{Erig}
    S (q,\nu) = \delta(\nu)
\,.
\end{equation}
Quantum-mechanical ground-state oscillations will be neglected.
This model serves to simulate normalisation scans.
The need for such simulations will become apparent in section \ref{RE}.

\subsection{Glass model} \label{MT}

The {\it glass} model describes an isotropic assembly of harmonic oscillators.
The ideal scattering law $S(q,\nu)$ is calculated by
explicit Fourier transform of
\begin{equation}\label{ESqt}
    S(q,t) = {\rm e}^{-2W(q,0)} {\rm e}^{2W(q,t)}
\,.
\end{equation}
In the high-temperature limit the exponents are given by
\begin{equation}\label{EWqt}
   2W(q,t) = {\hbar^2 q^2\over 6 M k_{\rm B} T}
   \int\!{\rm d}\nu\, {\rm e}^{-i2\pi\nu t} 
   {\left({k_{\rm B}T\over h\nu}\right)}^2 g(|\nu|)
\,.
\end{equation}
where $T$ is the temperature of the sample
and $M$ the average mass of the atoms.
Since the sharp cut-off of the Debye-model
leads to overshots in the Fourier transform,
it is preferable to assume a smooth density of states,
\begin{equation}\label{Edos}
   g(\nu) = {9\nu^2\over{\nu_{\rm D}}^3}\, 
             \exp \left({-\left({9\pi\over16}\right)^{1/3}
                          \left({\nu\over{\nu_{\rm D}}}\right)^2}\right)
\,.
\end{equation}
The Debye frequency 
$   \nu_{\rm D} = {\left({3 n / 4\pi}\right)}^{1/3} c $
depends on the atomic density $n$ and the sound velocity $c$
which has to be calculated as an average 
$\langle c^{-3}\rangle^{-1/3}$ over the longitudinal and transverse modes.
For this model, 
the mean-square displacement can be calculated:
\begin{equation}\label{Emsd}
   {r_0}^2 = 2W(q,0)/q^2 = {\left({9\pi\over2}\right)}^{1/3}
       {k_{\rm B}T\over M {(2\pi\nu_{\rm D})}^2}
\,.
\end{equation}
The parameter set
\begin{equation}\label{Epgl}
\begin{array}{@{}lcl}
n &=& 10^{23}\mbox{ cm}^{-3}\,,\\*[1.5ex]
c &=& 1.2\mbox{ km/s}\,,\\*[1.5ex]
M &=& 7.1\mbox{ a.m.u., and}\\*[1.5ex]
T &=& 150\mbox{ K}
\end{array}
\end{equation}
models reasonably well an organic molecular or polymeric glass;
it leads to a displacement $r_0=0.3$~\AA\ 
and to a Debye frequency $\nu_{\rm D}=3.46$~THz.

\subsection{Liquid model} \label{MM}

The {\it liquid} model is defined by a simple mode-coupling model
\begin{equation}\label{Eeqm}
\begin{array}{@{}lcl}
    0 &=& \ddot{\phi}_x(t) + \eta_x \dot{\phi}_x(t) + {\Omega_x}^2 \phi_x(t)
\\*[2.2ex]
    && \displaystyle
       +  {\Omega_x}^2 \int_0^t\!{\rm d}t'\,m_x(\{\phi\},t-t')\dot{\phi}_x(t')
\end{array}
\end{equation}
where the subscript $x$ denotes either density correlations
around the structure factor maximum ($x=0$),
or tagged-particle correlations 
at different wavenumbers ($x=q$).
The characteristic frequencies $\Omega_x$ set the time scale;
the friction term $\eta_x \dot{\phi}_x$ stands for fast force fluctuations
that have no influence on the long-time dynamics.

With the initial conditions
\begin{equation}\label{Eini}
   \phi_x(0) = 1\,,\quad\dot{\phi}_x(0)=0
\end{equation}
and the memory kernel of the $F_{12}$ model \cite{Got91,Got84},
\begin{equation}\label{Ecke}
   m_0(\{\phi\},t) = v_1 \phi_0(t) + v_2{\phi_0(t)}^2
\,,
\end{equation}
the collective dynamics $\phi_0(t)$ is fully determined
by the coupling coefficients $v_1(T)$, $v_2(T)$.
The tagged-particle correlators $\phi_{q}$, on the other hand,
are driven by $\phi_0$.
The simplest, bilinear coupling
\begin{equation}\label{Eske}
   m_{q}(\{\phi\},t) = v_{q} \phi_0(t) \phi_{q}(t)
\end{equation}
is designated as Sj\"ogren model \cite{Sjo86}.
The incoherent scattering law $S(q,\nu)$ is obtained 
by Fourier transform of $\phi_{q}(t)$.

The most striking prediction of mode-coupling theory is probably
the existence of an intermediate scaling regime
 between $\alpha$ relaxation and microscopic vibrations,
where all time correlation functions $\phi_x$ slow down 
towards a plateau $f_x$ \cite{x50}.
Around this plateau, they factorize as
\begin{equation}\label{Efac}
   \phi_x(t)-f_x = h_x g_\lambda(t/t_\sigma)
\,.
\end{equation}
The shape of the universal scaling function $g_\lambda$ depends on just
one global parameter~$\lambda$.
Further predictions are made for the critical temperature dependence
of $h_x$ and~$t_\sigma$.
Many neutron scattering experiments 
\cite{WuKB93,KnMF88,DoCP90a,FrZR91,WuHL94,ToPD96,RuET97,MeWP98a,WuSH98,WuOG00a}
have been undertaken
to test these predictions.
However, 
the asymptotic law~(\ref{Efac}) holds only in a restricted frequency range,
and therefore it cannot be used as input to a multiple-scattering calculation.

In the last couple of years it became possible to calculate
the full evolution of $\phi_x(t)$
very efficiently and to arbitrarily long times
by explicit integration in the time domain \cite{Sin95,Got96}.
In cases where the asymptotic regime is not reached
numeric solutions of schematic mode-coupling models 
have been used to fit experimental data \cite{AlKr95,KrAK97,FrGM97,RuET99}.
In a most recent example
data from incoherent neutron scattering \cite{WuOG00a},
depolarized light scattering \cite{WuOG00a,DuLC94} 
and dielectric spectroscopy \cite{ScLB99}
on glass-forming propylene carbonate 
have been analysed first in terms of scaling \cite{WuOG00a}
and then by integration of the $F_{12}$--Sj\"ogren model,
where the different observables
were all governed
by one and the same density correlator $\phi_0(t)$ \cite{GoVo00}.
Results from these fits will now be used to construct a realistic
$S(q,\nu)$ as input to a multiple-scattering simulation.

We arbitrarily select the 220 K data
which could be fitted with the following 
set of parameters \cite{GoVo00,x45,x53}:
\begin{equation}\label{Epar}
\begin{array}{@{}lcl}
\Omega_0 &=& 1000 \mbox{ GHz}                   \,,\\*[.3ex]
\Omega_q &=& q \cdot 224 \mbox{ GHz / \AA}^{-1} \,,\\*[.3ex]
\eta_0   &=& 0                                  \,,\\*[.3ex]
\eta_q   &=& 350 \mbox{ GHz}                    \,,\\*[.3ex]
v_1      &=& 0.83                               \,,\\*[.3ex]
v_2      &=& 1.66                               \,.
\end{array}
\end{equation}
Deviating from Ref.~\cite{GoVo00}, the $q$-dependent vertices in the
Sj\"ogren coupling~(\ref{Eske}) are determined from
\begin{equation}\label{Egof}
    1 / (v_q f_0) = 1 - \exp ( - {r_0}^2 q^2 )
\end{equation}
with $r_0=0.546$~\AA,
which satisfies the physical requirements
$1\!-\!f_q\sim q^2$ and $h_q\sim q^2$ for $q\to0$ as well as
$f_q\to0$ for $q\to\infty$ \cite{x54}.

\section{Monte-Carlo simulation}


\subsection{Algorithm} \label{SA}

The multiple-scattering simulation consists essentially of 
a Monte-Carlo integration over many neutron trajectories.
The program basically follows the well documented {\sc Mscat} algorithm
\cite{BiYM72,Cop74,CoVW86}.
All restrictions on storage size could be lifted;
the quasielastic scattering law was stored on logarithmic $q$ and $\nu$ grids
with about $40\times240$ entries.
Runs with $10^4$ to $10^6$ neutrons on a medium-size workstation
took between less than a minute and several hours.

Each neutron is initialized with an energy $E_0$
and a direction ${\bf\hat k}_0$ along the incident beam.
Since we are not interested in instrumental resolution effects,
the option of choosing $E_0$ and ${\bf\hat k}_0$ 
from finite distributions is not used.
Next, the impact point ${\bf r}_0$ 
on the sample surface is chosen at random,
and the length $l({\bf r}_0,{\bf\hat k}_0)$
of a trajectory straight across the sample is calculated.
Given the total scattering cross section density $\Sigma(E_0)$,
the neutron will be scattered somewhere within the sample 
with a probability
$p_0=\exp(-\Sigma(E_0)l({\bf r}_0,{\bf\hat k}_0))$.
With a probability $1-p_0$, 
the neutron will traverse the sample without interaction;
absorption shall not be considered.
At this point, 
the algorithm forces all neutrons to be scattered within the sample,
assigning them as a weight $w_0$ the survival probability $p_0$.
A collision point ${\bf r}_1$ 
is chosen at a distance $l$ from ${\bf r}_0$ 
with a probability proportional to ${\rm d}\exp(-\Sigma(E_0)l)/{\rm d}l$,
and a new energy $E_1$ and direction ${\bf\hat k}_1$ are selected
according to the ideal scattering law $S(q,\nu)$.
Then,
the distance $l({\bf r}_1,{\bf\hat k}_1)$ to be travelled
upon leaving the sample is calculated,
the neutron is assigned a new weight $w_1=w_0 p_1$,
and the whole procedure is iterated.

For each collision $i=1,2,\ldots$, 
the contribution of the neutron to the 
scattering score $S_{(i)}(2\theta,\nu)$
is evaluated for all detector angles and for all energy channels.
The weight of each contribution is a product of (i) the weight $w_{i-1}$,
(ii) the scattering law that brings the neutron from its previous state
into the segment $q,\nu$,
and (iii) the probability of reaching the detector
without further collsions.

With each collision the neutron looses weight.
Following its trajectory too far would make the simulation inefficient.
Therefore, when the weight $w_i$ falls below a predefined threshold $w_{\rm c}$,
the neutron's fate is determined by a Russian roulette:
with a probability $1/2$ its weight is doubled,
otherwise the trajectory has come to an end.

\subsection{Setup} \label{SS}

Samples have most often the form of a hollow cylinder 
(with its axis perpendicular to the scattering plane) 
or of a slab
(with its normal vector in the scattering plane).
Here we choose the cylindrical geometry 
which is preferred in experiments
because it is easy to prepare and to seal,
and at the same time it keeps
 self-shielding and multiple-scattering effects
rather isotropic \cite{Wut99,Wut99E}.

In slabs
flight paths become very long
when neutrons are scattered into the sample plane.
For scattering angles around the mounting angle of the slab
so many neutrons are lost by absorption or multiple scattering
that no meaningful signal is measured.
Outside this region 
multiple-scattering effects are expected not
to depend critically on the sample geometry.
In particular, we expect that our low-$q$ results
hold qualitatively for slabs as well as for cylindrical samples.

To proceed, our cylinder has a height of 50 mm and an outer diameter of 30 mm,
and it is fully illuminated by the incident beam.
The simulation does not attempt to describe resolution effects of 
the secondary spectrometer;
therefore the detectors are placed at infinite distance from the sample.

The bound cross section density is 
$\Sigma_0 = 80$~barn $\times$ $5\cdot10^{22}$~cm$^{-3} = 0.4$~mm$^{-1}$,
which is a typical value for hydrogen-rich organic materials.
In the low-temperature limit of a rigid scatterer, 
$\Sigma_0$ is equal to the total cross section density $\Sigma(E_0)$;
at higher temperatures, $\Sigma(E_0)$ is a bit bigger.
The absolute scattering power of the sample 
depends on the thickness $b$ of the tubular layer.
In practice one characterises the sample thickness by the transmission
of a collimated beam,
\begin{equation}\label{Etra}
    T_{\rm coll} = \exp (-\Sigma(E_0)2b)
\,.
\end{equation}
Samples with $T_{\rm coll}\simeq0.9$ are generally regarded as
a good compromise between the conflicting requirements
of high single-scattering and low multiple-scattering rates.
According to often heard folklore,
a sample with 90\,\% transmission is a 10\,\% scatterer,
and therefore about 10\,\% of the scattered neutrons will 
undergo a second collision.
As explained in Ref.~\cite{Wut99E} this is not generally true:
in a tubular sample 
one needs a transmission of 96\,\% 
(properly measured with a collimated beam)
in order to obtain a 6\,\% scatterer (with reference to the full beam),
in which about 10\,\% of the scattered neutrons be scattered a second time.

For the present work, samples of different thickness have been studied.
In order to highlight the effects of multiple scattering,
most results will be shown for a relatively thick sample 
with $b=0.3$~mm, corresponding to
a transmission $T_{\rm coll}=0.79$.
In Fig.~\ref{Feno}, elastic scattering will be discussed as function of~$b$.

\begin{figure}
\epsfxsize=81mm\centerline{\epsffile{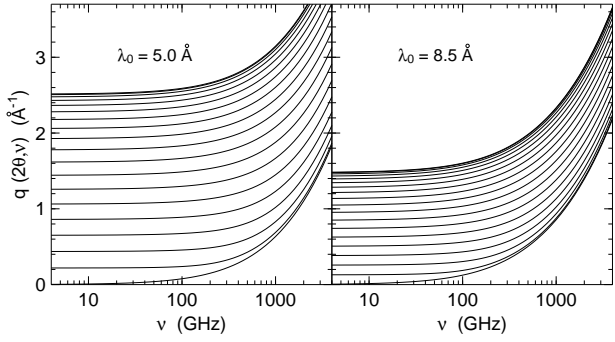}}~\\
\caption 
{Dynamic window for inelastic neutron scattering
with the two incident wavelengths $\lambda_0=5.0$~\AA\ 
and $\lambda_0=8.5$~\AA\ used in this study.
The lines show $q(2\theta,\nu)$ 
for scattering angles from $2\theta=0^\circ$ (bottom) to $180^\circ$ (top)
in steps of $10^\circ$.}
\label{Fdyw}
\end{figure}

As in a real experiment, the incident neutron wavelength 
has been adapted to the physics under study:
A wavelength $\lambda_0=5.0$~\AA\ has been chosen
for the scattering from phonons in the {\it glass} model,
and a longer wavelength $\lambda_0=8.5$~\AA\ for
the investigation of fast relaxation in the {\it liquid} model.
Fig.~\ref{Fdyw} shows the dynamic windows 
that are accessible under these conditions.

On output, 
the simulation yields the scattering contributions 
at constant detector positions $2\theta$.
Just as experimental data, 
these $S_{(i)}(2\theta,\nu)$ must be interpolated
to constant wavenumbers~$q$ before they can be physically interpreted.
The interpolation $q\to2\theta\to q$ is also performed on the
{\it ideal} scattering law which therefore may slightly deviate
from the model law $S(q,\nu)$ used as input to the simulation.

\section{Results}

\subsection{Elastic scattering and normalisation} \label{RE}

\begin{figure}
\epsfxsize=60mm\centerline{\epsffile{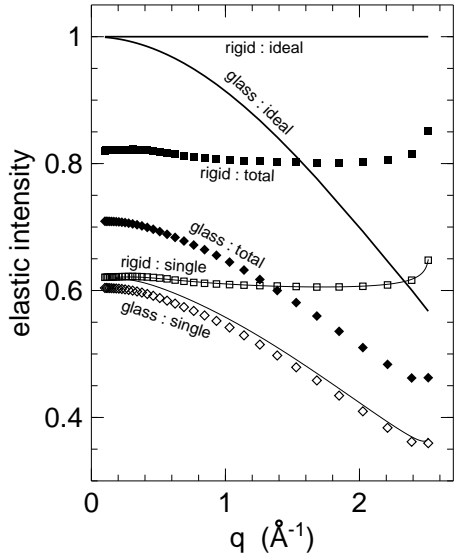}}~\\
\caption 
{Elastic intensity $I(q)$ from
simulated scattering experiments.
The incident neutrons have a wavelength $\lambda_0=5.0$~\AA;
the sample is tubular with a transmission $T_{\rm coll}=0.79$,
as described in Sect.~\protect\ref{SS}.
The ideal scattering law, used on input, 
is given by the {\it rigid} model (Sect.~\protect\ref{MR})
and the {\it glass} model (Sect.~\protect\ref{MT}).
The thick lines show the amplitude of the elastic part of
the ideal scattering law;
open symbols represent single scattering,
and full symbols stand for the sum of single and multiple scattering.
The thin lines have been calculated as product of the ideal scattering law
with the self-shielding coefficient $A(q)$
for elastic scattering.}
\label{Fela}
\end{figure}

\begin{figure}
\epsfxsize=78mm\centerline{\epsffile{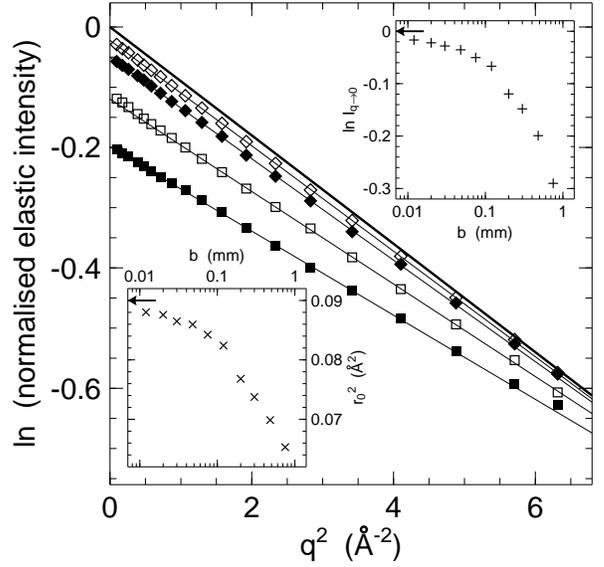}}~\\
\caption 
{Elastic intensity of the {\it glass} model,
normalized to the {\it rigid} model, 
shown as $\ln I(q)$ {\it vs.} $q^2$,
samples of different thickness
(from top to bottom: $b=0.02$, 0.075, 0.2, 0.48 mm,
corresponding to transmissions from $T_{\rm coll}=0.984$ to $0.68$).
The thick line shows the Gaussian elastic intensity given on input;
the thin lines are Gaussian fits $I(q)=I_0\exp(-{r_0}^2 q^2)$ 
to an intermediate-$q$ region.
The insets shows the so-obtained parameters $\ln I_0$ and 
${r_0}^2$ as function of $b$.
For thin samples, they converge quite slowly towards the
ideal values $I_0=1$ and ${r_0}^2=0.09$~\AA$^2$ (arrows).}
\label{Feno}
\end{figure}

Results from selected simulations are presented 
in Figures~\ref{Fela}--\ref{Fult}.
The analysis starts with Fig.~\ref{Fela}
which shows the elastic scattering from the {\it rigid}
and the {\it glass} model.
As in most of the following figures,
the ideal scattering law of the model is compared 
to the total scattering registered in the simulated experiment.
Additionally, Fig.~\ref{Fela} shows
which part of the total scattering is due to single scattering.

For the {\it rigid} model
the single-scattering intensity $I_{(1)}(q)$
is equal to the self-shielding coefficient $A(2\theta(q))$.
This presents an important test of the Monte-Carlo code
(and actually led to discovering an error in the
determination of $A(2\theta)$ \cite{Wut99E}).
In the {\it glass} model,
the possibility of inelastic scattering augments 
the cross-section density $\Sigma(E_0)>\Sigma_0$,
and therefore $I_{(1)}(q)$ is somewhat smaller than 
the product of $A(2\theta(q))$ and the ideal elastic intensity
$I_{\rm ideal}(q)=\exp(-{r_0}^2q^2)$.

The multiple-scattering contribution is almost isotropic.
For a rigid scatterer in our relatively thick standard geometry
(with $T_{\rm coll}=0.79$)
it varies by only $\pm2$\,\% around the average value $I_{\rm multi}=0.20$.
In the glass the elastic multiple scattering sinks by about one half
to $I_{\rm multi}=0.10$ with a wavenumber-dependent variations 
still of the order of~$\pm2$\,\%.
The total scattering, 
obtained as the sum of single and multiple scattering,
remains for all wavenumbers below $I_{\rm ideal}(q)$.
Even in the limit $q\to0$, 
where the incoherent scattering law necessarily goes to $I_{\rm ideal}(q)\to1$
the simulated signal remains smaller than~1.
This intensity defect has been observed in many experiments 
(clearly shown e.g. in \cite{FrRP88,FeDP93,FrRi93,MeDS97}),
and simulations \cite{Bee88} have confirmed
multiple scattering as its likely cause.

Multiple-scattering effects in the {\it rigid} model
bring us to the problem of normalisation:
while Monte-Carlo simulations are able to produce $S_{\rm total}(q,\nu)$
in {\it absolute} units,
experiments are not.
In experiments, 
the scattering law is always measured relative to 
that of a well-known incoherent standard scatterer.
Usually, this standard scatterer is vanadium.
If the sample to be studied is itself an incoherent scatterer,
a better choice is normalisation 
to its own low-temperature elastic response.
In both cases, the normalisation scan
is well represented by our {\it rigid} model.

As Fig.~\ref{Fela} demonstrates
normalisation of the {\it glass} to the {\it rigid} model
reduces the $q\to0$ intensity defect by about a factor~2.
Thus, 
{\it multiple-scattering simulations will never become quantitatively useful
without simulating the normalisation scan as well}.
Consequently, all simulated data presented in the remainder of this paper
are normalized to the {\it rigid} model simulation.

Fig.~\ref{Feno} shows normalized elastic intensities
of the {\it glass} model for samples of different thickness~$b$.
In the common representation $\ln I(q)$ {\it vs.} $q^2$,
Gaussians
\begin{equation}\label{EIq0}
    I(q) = I_0 \exp (- {r_0}^2 q^2)
\end{equation}
appear as straight lines.
The ideal scattering law is Gaussian by construction,
with $I_0=1$ and $r_0=0.30$~\AA.
As anticipated, the simulations yield intersections $I_0<1$.
The question is \cite{FrRi93}
whether in this situation fits with Eq.~(\ref{EIq0})
can still be used to extract a meaningful displacement~$r_0$.
The inset of Fig.~\ref{Feno} gives an affirmative answer:
for samples with $T_{\rm coll}\gtrsim0.8$,
$r_0$ will be underestimated by less than 10\%.

\subsection{Phonons} \label{RP}

The inelastic scattering from the {\it glass} model is quite weak.
Very long runs 
are necessary before the simulated scattering law can be analysed.
Figure~\ref{Fptt} shows results from simulations with 
$10^6$ neutrons.
In the upper frame
simulated data are plotted as obtained at constant detector angles;
in the lower frame they have been interpolated to constant wavenumbers.

At small angles the interrelation between $2\theta$, $q$  and $\nu$ 
causes the small-angle scattering law $S(2\theta,\nu)$
to attain a maximum at between 2 and 3 THz
whereas $S(q,\nu)$ decreases monotonically for any given~$q$.
Similar anomalies affect also the multiple scattering.
Therefore, observations in this part of the dynamic window
are likely to depend on the incident neutron wavelength \cite{x49}.

The present work will concentrate on the more generic effects of
multiple scattering at lower frequencies where 
a given scattering angle corresponds to an almost constant wavenumber.
In this region the inelastic scattering from the glass model
is essentially constant, $S(q,\nu)=J_q$.
Since the simulations have been performed on a logarithmic frequency grid,
best accuracy is achieved by calculating $J_q$
as a logarithmic average
\begin{equation}\label{Eiav}
   J_q = \int_{\nu_1}^{\nu_2}\!{\rm d}\ln\nu\,S(q,\nu) / 
         \int_{\nu_1}^{\nu_2}\!{\rm d}\ln\nu
\,.
\end{equation}
With $\nu_1=10$ GHz to $\nu_2=100$~GHz we concentrate on a range
where the curves $q(2\theta,\nu)$ vs $\nu$ are essentially flat 
[Fig.~\ref{Fdyw}].

The $q$ dependence of $J_q$ is shown in Figure~\ref{Fipi}.
In the $\nu\to0$ limit 
\begin{equation}\label{EIqI}
   J_q = \int\!{\rm d}t\, [ S(q,t) - S(q,\infty) ]
\end{equation}
one can develop Eqs.\ (\ref{ESqt}) and~(\ref{EWqt}) into
\begin{equation}\label{EIq2}
   J_q = \left({3\over4\pi}\right)^{1/3} {{r_0}^2\over\nu_{\rm D}} q^2 
          + {\cal O}(q^4)
\,.
\end{equation}
This motivates fits of the simulated intensity with a polynomial in~$q^2$,
\begin{equation}\label{Efpo}
   J_q \simeq A + B q^2 + C q^4
\,.
\end{equation}
For the ideal scattering law, 
one has $A=0$, and the coefficient $B$ agrees within 2\,\% 
with the expectation from Eq.~(\ref{EIq2}).
For the simulated scattering law, 
we find a considerable base line $A_{\rm tot}$,
and a coefficient $B_{\rm tot}\simeq0.75B$.

Sometimes a frequency-dependent version of Eq.~(\ref{Efpo})
is used for data analysis \cite{CuDo90,SeDo96}.
While multiple scattering is made responsible for $A_{\rm tot}(\nu)$
and $C_{\rm tot}(\nu)q^4$ is attributed to multi-phonon processes,
the $B_{\rm tot}(\nu)q^2$ is taken as an approximation
to the $q\to0$ limit of the ideal scattering law.
As we have seen, for our model (with $T_{\rm coll}=0.79$)
this ansatz underestimates $B(\nu)$ by about 25\,\%.
One can however expect that this error affects more the absolute
intensity scale than the frequency dependence of $S(q,\nu)/q^2$.

\subsection{Quasielastic spectra} \label{RQ}

The nontrivial features of quasielastic spectra
are visualized best after converting them to susceptibilities
\begin{equation}\label{Etriv}
\chi''_q(\nu) = S(q,\nu) / n(\nu)
\end{equation}
with the Bose factor $n(\nu)= {(\exp(h\nu/k_{\rm B}T)-1)}^{-1}$.
Figure~\ref{Fsus} shows the ideal and the simulated susceptibility 
of the {\it liquid} model at different wavenumbers.
We see a wavenumber-dependent $\alpha$ peak at low frequencies,
the scaling region of fast relaxation around the minimum at 60 GHz, 
and a vibrational peak a bit below 
the model's fundamental frequency $\Omega_0=1$~THz.

At large wavenumbers, this scenario is qualitatively reproduced 
in the simulated experiment,
although the spectral distribution is significantly distorted by
multiple scattering.
The simulated susceptibilities even cross the input curves:
in the phonon range, {\it more} neutrons arrive than expected from
the ideal scattering law,
similar to what was found for the {\it glass} model [Fig.~\ref{Fptt}].

At small wavenumbers, 
multiple scattering changes the susceptibilities even qualitatively:
in addition to the $\alpha$ peak of the ideal scattering law
the simulated small-angle data possess another peak,
which is entirely due to multiple large-angle scattering.
Around this peak, 
{\it multiple scattering is up to two orders of magnitude 
stronger than single scattering.
Such anomalies can arise as soon as the ideal scattering law
has a pronounced wavenumber dependence.}

For a quantitative analysis, the $\alpha$ peaks have been fitted with the
Fourier transform \cite{DiWB85,ChSt91} of the Kohlrausch stretched exponential
\begin{equation}\label{Ekww}
   \Phi_q(t) = A_q \exp(-(t/\tau_q)^{\beta_q})
\,.
\end{equation}
The wavenumber-dependent fit parameters are reported in Fig.~\ref{Fapa}.
Instead of $\tau_q$, the mean relaxation time
\begin{equation}\label{Etav}
   \langle\tau_q\rangle = \int_0^\infty\!{\rm d}t\, {\Phi_q(t)\over\Phi_q(0)}
      = {\tau_q\over\beta}\Gamma({1\over\beta})
\end{equation}
is shown because it couples less strongly to $\beta_q$.
The representation as $q^2\langle\tau_q\rangle$
anticipates an overall wavenumber dependence 
$\langle\tau_q\rangle\propto q^{-2}$,
which is well fulfilled in the small-$q$ limit
where tagged-particle motion can be described 
as simple diffusion \cite{WuCR96b,BoYi80}.
Even for the ideal scattering law the fit parameters show random fluctuations,
which are due to trivial inaccuracies in interpolating
from $q$ to $2\theta$ and back.
The fluctuations are particularly strong in $\beta_q$
because only the very beginning ($\nu\!<\!2.5\,\nu_{\rm p}$)
of the high-frequency wing was fitted.

Nevertheless we can read off with certainty
that multiple scattering affects the line shape and the time constant
much less than the amplitude.
Multiple-scattering effects are most pronounced at intermediate wavenumbers:
at small wavenumbers the spurious $\alpha$ peak from multiple scattering is
so far away that it distorts no longer the top of the single-scattering
$\alpha$ peak.

In Figures~\ref{Fbmi}--\ref{Fhqm}
we shall analyse the scaling behaviour of the fast relaxation.
Around the minimum of $\chi''(q,\nu)$
the factorisation property (\ref{Efac}) implies that
all susceptibilities can be rescaled onto a master curve
\begin{equation}\label{Esfa}
    \hat\chi_q''(\nu) = \chi''(q,\nu) / h_q
\,.
\end{equation}
The amplitudes are determined from the simulated $\chi''(q,\nu)$
by a least-squares match of neighbouring $q$ cuts,
just as one would do in the analysis of experimental data
\cite{WuKB93,WuSH98,WuOG00a}.

Figure~\ref{Fbmi} shows the $\hat\chi_q''(\nu)$.
Around and above the susceptibility minimum,
the simulated data fall quite well onto each other.
At lower frequencies, 
the cross-over towards the $\alpha$ peak 
leads to wavenumber-dependent multiple-scattering effects
that cause small but systematic violations of the factorisation.
Here again, multiple-scattering effects are least at large angles.

Therefore, in Fig.~\ref{Ffmi} the analysis is restricted 
to wavenumbers above $1.0$~\AA$^{-1}$.
In this range ideal and simulated susceptibilities 
are $q$ independent
over a frequency range of more than a decade around the minimum.
The average ${\langle{\hat\chi_q''(\nu)}\rangle}_q$ 
are fitted by the scaling function $g_\lambda(\hat\nu)$ \cite{Got90}.
As in many real experiments the fits work only for frequencies below
the minimum.
The ideal scattering law is described by $\lambda=0.73$.
This value differs considerably from the parameter 0.775 used as input
to the model construction [Eq.~(\ref{Epar})],
which is not unexpected in a physical situation
in which the asymptotic regime described by Eq.~(\ref{Efac}) 
is not fully reached.
Nevertheless, as discussed in Ref.~\cite{GoVo00},
the asymptotic formul\ae\ give an adequate qualitative description
of the experimentally accessible dynamics.
A fortiori, fits with $g_\lambda(\hat\nu)$
remain useful for communicating experimental results
and for comparing results from different sources~\cite{WuOG00a}.

In this sense, the simulated data in Fig.~\ref{Ffmi}b
shall also be fitted with the asymptotic scaling function.
One finds almost exactly the same $\lambda$ as from the fit to the
ideal susceptibility.
Although this accord may be to some degree coincidental,
it shows that 
large-angle susceptibilities in the fast relaxation regime
are not easily distorted by multiple scattering.
On the other hand, the minimum position $\nu_\sigma$ is shifted
from 63 to 50~GHz. 

Figure~\ref{Fhqm} shows the amplitude $h_q$.
For the ideal scattering law $h_q$ is proportional to $1-f_q$,
with a Gaussian $f_q$, 
as expected from the model's construction.
For the simulated data, 
the wavenumber dependence of $h_q$ is smeared out considerably.
The small-wavenumber limit $h_q\propto q^2$ sits now on top of
a huge constant term. 
Towards larger wavenumbers, the $h_q$ increase less than in the ideal case.
In the range $0.8$ \AA$^{-1}\lesssim1.6$~\AA$^{-1}$
this leads to a nearly perfect though physically meaningless
linear behaviour $h_q\propto q$
(similarly, one could draw a line $J_q\propto q$ through the phonon data 
of Fig.~\ref{Fipi}).
Such a linearity has been observed in several experimental studies
\cite{KiBD92,x48}
--- most recently in exactly the same wavenumber range for 
propylene carbonate \cite{WuOG00a}.
It has been suspected from the beginning that 
this behaviour and in particular the deviations from the physical
small-$q$ limit $h_q\sim q^2$ are due to multiple scattering.
The present results show that this explanation is consistent and plausible.

\subsection{Scattering angles} \label{RA}

The Monte-Carlo simulation not only yields the total scattering law
$S(2\theta,\nu)$ and its partials $S_{(i)}(2\theta,\nu)$ ---
with simple extensions the code can also be used to
generate additional information
that is not accessible in experiments.
For instance it is possible to score conditional probabilities
that describe which single-scattering events $\{2\theta_i,\nu_i\}$
contribute to
the multiple-scattering counts registered in a given channel $2\theta,\nu$.
Here we shall consider the simplest case:
elastic double-scattering from the {\it rigid} model.
Given a double-scattered neutron that arrives at a detector angle $2\theta$,
we ask for the probabilities $f_i(2\theta_i|2\theta)$
that in the $i$-th collision ($i=1,2$) the neutron has been scattered
by an angle $2\theta_i$.

A simulation with some $10^4$ neutrons confirms $f_1=f_2$.
This was expected from symmetry and 
allows us to improve the statistics by calculating an
average $f=(f_1+f_2)/2$.
Figure~\ref{Fadi} shows $f(2\theta'|2\theta)$ as function of 
the single-scattering angle~$2\theta'$.
Surprisingly, this function shows no siginificant dependence on 
the total scattering angle~$2\theta$.
For any~$2\theta$, 
it is an almost triangular function of~$2\theta'$, 
except around the maximum at $2\theta'=90^\circ$
where it is even somewhat sharper.
This is the joined effect of two causes:
The solid angle accessible for a given interval in $2\theta'$
is proportional to $\sin2\theta'$.
And for scattering angles
around $90^\circ$ there is a chance
that the flight path between the two collisions is about perpendicular
to the scattering plane, 
and thus parallel to the symmetry axis of the tubular sample.
In this case, neutrons have to travel a very long path before
leaving the sample, and therefore they will almost certainly be available
for a second scattering process, 
thereby enhancing their contribution to~$f(2\theta'|2\theta)$.

\section{Conclusion}

Starting with {\it elastic} scattering,
we have reconfirmed that multiple scattering leads to a pronounced
intensity defect in $I(q\!\to\!0)$,
as regularly observed in back-scattering measurements.
The strong effects of multiple scattering in the {\it rigid} model
make clear that any correction of experimental data
must start with correcting the normalisation scan.

With increasing temperature 
(passing to the {\it glass} model)
part of the neutrons goes in inelastic channels;
the elastic scattering probability $I_{\rm ideal}(q)$
becomes $q$ dependent and diminishes on average.
This leads to a strong decrease of the elastic-elastic multiple-scattering
but does not change its angular distribution which remains
almost isotropic.
Even for a rather thick scatterer 
the $q$ dependence of the total elastic intensity 
remains close to the input Gaussian.
This can be seen as support for the optimistic view \cite{Zor97b}
according to which it is not impossible, after appropriate corrections,
to extract additional information 
from subtle features of a non-Gaussian elastic intensity.

Passing to {\it inelastic} scattering,
it has been known for long 
that multiple scattering distorts more the wavenumber dependence of $S(q,\nu)$
than its frequency dependence.
The reason is quite simple:
in a typical solid, as represented by our {\it glass} model,
and for typical neutron wavelengths, as chosen in a time-of-flight experiment,
the Debye-Waller factor is not too different from 1,
which means that most scattering events are elastic.
Under this condition,
a double-scattering event registered in an inelastic channel
is much more likely to stem from an elastic-inelastic or inelastic-elastic
history than from a sequence of two inelastic collisions.
Since the amplitude initially goes with $J_q\propto q^2$
it follows that multiple scattering has its worst effects
on small-angle measurements.
These insights are fully confirmed by the present simulation.
It is shown that multiple scattering can lead 
to an appealling yet unphysical $J_q\propto q$ dependence.
It is emphasized that high frequencies give rise to additional difficulties
because constant-angle detectors measure at frequency-dependent wavenumbers
$q(2\theta,\nu)$.

Taking advantage of recent progress in handling mode-coupling equations
it was possible to construct a {\it liquid} model,
which not only describes relaxational dynamics but comprises at least
schematically also the vibrational spectrum so that it is defined
in the entire $q,\nu$ plane.
Simulations on this model show at least one bizarre effect ---
the shadow $\alpha$ peak in Fig.~\ref{Fsus} ---
but as a whole they are reassuring:
as in the glass, multiple scattering distorts much more the
wavenumber dependence than the frequency dependence of~$S(q,\nu)$.
The elastic line is quasielastically broadened,
but one can still argue that (almost elastic)-(not so elastic) histories
are much more probable than (not so elastic)-(not so elastic) sequences.
As in the glass, the frequency distribution suffers least 
at the largest scattering angles.
At these angles
the line shape of the $\alpha$ peak can be determined with good precision;
around the susceptibility minimum the line shape of fast $\beta$ relaxation
is not at all distorted by multiple scattering.
The position of the minimum is shifted by a small amount
which however is not completely negligible 
when compared to the degree of agreement reached
between neutron scattering and 
fundamentally different experimental techniques 
(Fig.~14 of Ref.~\cite{WuOG00a}).
The amplitude~$h_q$ of the susceptibility minimum behaves very similar 
to the phonon intensity~$J_q$:
the asymptotic $q^2$ dependence sits on top 
of an isotropic multiple-scattering contribution,
leading to an apparent $h_q\propto q$ behaviour 
in the experimentally relevant wavenumber range.
This is a central result of the present work
because it answers a question 
that had been pending for many years \cite{KiBD92}
and still remained open in the extensive data analysis of
Refs.~\cite{WuOG00a,GoVo00}.

On a technical level,
the present work illustrates 
that the main effort in studying multiple-scattering goes into
the formulation of dynamic models that 
are physical, tractable and complete 
(covering a wide $q,\nu$ region, 
thereby also guaranteeing correct normalisation).
The simulation itself is a routine operation,
once one has adapted the Monte-Carlo code to one's personal needs.
In this situation, 
the results of the angular scoring [Sect.~\ref{RA}, Fig.~\ref{Fadi}]
open a new perspective:
Only very few multiple-scattering sequences involve
extreme scattering angles 
that are not covered in a multi-detector experiment.
A vast majority of all multiple-scattering events depends only 
on the scattering law at intermediate angles.
Therefore, 
it seems possible to construct a sufficiently complete dynamic
model from the measured data alone.
This supports the ``pragmatic approach''
mentioned in the introduction.

The present results are expected to apply qualitatively
for any noncrystalline system.
Whenever $S(q,\nu)$ factorises into a
$q$-dependent amplitude and 
an essentially $q$-independent function of frequency,
the frequency distribution will suffer much less from multiple
scattering than the amplitude.
On the other hand,
when the scattering law has $q$-dependent maxima
multiple scattering may be lead to spurious peaks, especially at small angles.
In such situations,
simulations of more specific models must be undertaken.

\section*{Acknowledgments}

I thank Matthias Fuchs, Wolfgang G\"otze and Thomas Voigtmann for help 
with the mode-coupling model,
and Wolfgang Doster and Andreas Meyer for a critical reading of the manuscript.


\def\dirbib{}
\def\BIBsubm{submitted}
\def\BIBinpr{in press}
\bibliographystyle{\dirbib switch}
\bibliography{\dirbib jw1}

\begin{thebibliography}{10}
\bibitem{x53}For consistency with lower temperatures, the original fits of
  Ref.~\protect\cite{GoVo00} also comprise a hopping term. With a strength
  $\delta=7\cdot10^{-6}$ hopping does not influence the dynamics at 220~K and
  will therefore be neglected.
\bibitem{x54}It is an inherent weakness of the Sj\"ogren model that the fourth
  condition $h_q\to0$ for $q\to\infty$ cannot be satisfied simultaneously.
\bibitem{x49}This may explain (at least partially) the discrepancies shown in
  Fig.~2b of Ref.~\protect\cite{WuOG00a} where time-of-flight measurements with
  different incident wavelengths are compared.
\end{thebibliography}

\newpage
\narrowtext

\begin{figure}
\epsfxsize=72mm\centerline{\epsffile{mqm-05.ps.bb}}~\\
\caption 
{Inelastic scattering from the {\it glass} model,
shown at constant detector angles~$2\theta$ (upper frame)
and interpolated to constant wavenumbers~$q$ (lower frame).
The symbols show the simulated total scattering law;
the lines represent the ideal scattering law
(at the same~$2\theta$ or~$q$ and in the same order as the symbols).}
\label{Fptt}
\end{figure}

\begin{figure}
\epsfxsize=60mm\centerline{\epsffile{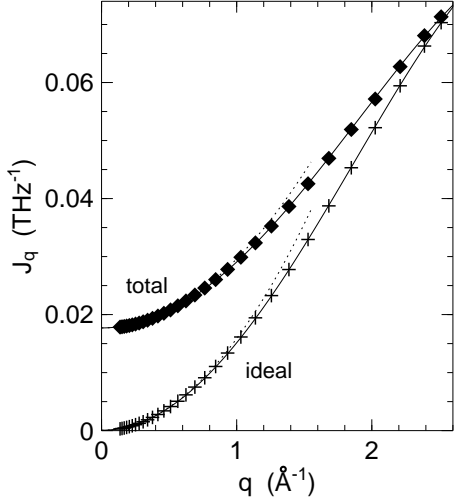}}~\\
\caption 
{Inelastic intensity from the {\it glass} model,
calculated as logarithmic average [Eq.~(\ref{Eiav})]
over the low-frequency region 10--100~GHz.
The full symbols show the normalized total scattering,
The plus signs represent the ideal scattering law.
Full lines are fits with a quadratic function in~$q^2$;
dotted lines show the same fits without $q^4$ contribution.
Above 1 \AA$^{-1}$, the total scattering could be described
by a simple $J_q\propto q$ dependence for which however there is no 
physical basis.}
\label{Fipi}
\end{figure}

\begin{figure}
\epsfxsize=90mm\centerline{\epsffile{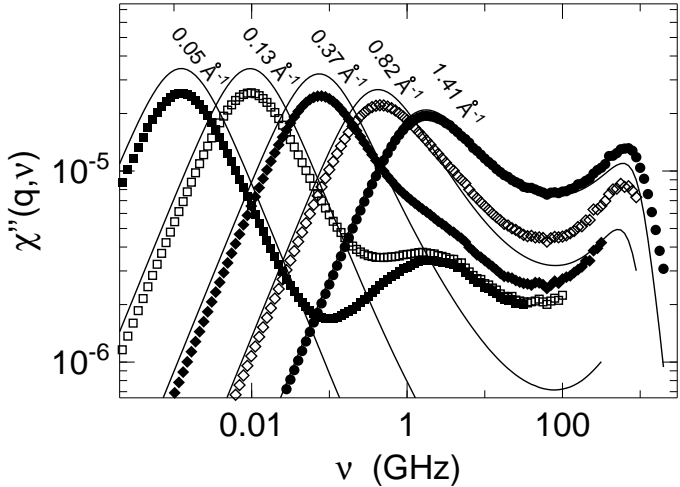}}~\\
\caption 
{Dynamic susceptibility of the {\it liquid} model,
simulated with an incident neutron wavelength $\lambda_0=8.5$~\AA.
Intensities are normalized to the {\it rigid} model.
Lines show the ideal scattering law,
symbols the simulated total scattering intensity.}
\label{Fsus}
\end{figure}

\begin{figure}
\epsfxsize=60mm\centerline{\epsffile{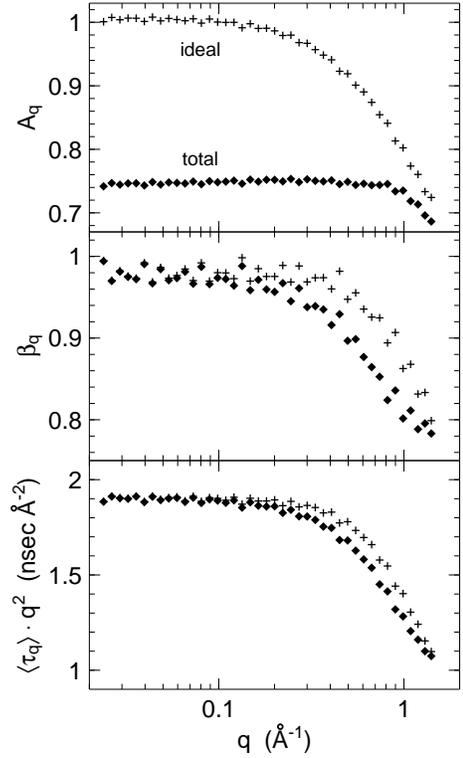}}~\\
\caption 
{Amplitude, stretching exponent and time constant 
from Kohlrausch fits of the $\alpha$ peak.
The different symbols refer to
the ideal ($+$) and simulated total ($\blacklozenge$) susceptibilities.
The time constants $\langle\tau_q\rangle$ have been multiplied with $q^2$.
Note that realistic experiments will only cover wavenumbers above about 
0.1~\AA$^{-1}$.}
\label{Fapa}
\end{figure}

\begin{figure}
\epsfxsize=78mm\centerline{\epsffile{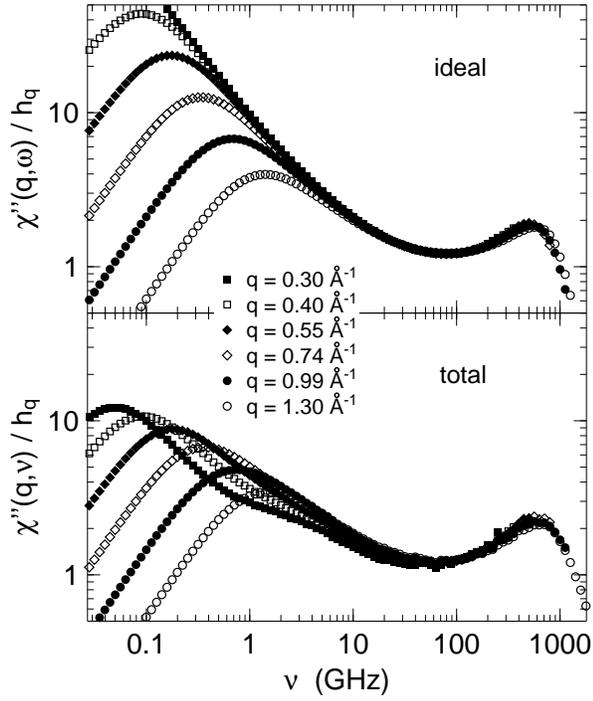}}~\\
\caption 
{Dynamic susceptibility as in Fig.~\protect\ref{Fsus},
rescaled with a $q$-dependent amplitude~$h_q$
according to the factorisation~(\protect\ref{Esfa}).
At small wavenumbers multiple scattering distorts in particular
the intensity ratio of $\alpha$ relaxation vs fast relaxation.
This can also be seen by comparing the amplitudes $A_q$ 
[Fig.~\protect\ref{Fapa}] and $h_q$ [Fig.~\protect\ref{Fhqm}].}
\label{Fbmi}
\end{figure}

\begin{figure}
\epsfxsize=78mm\centerline{\epsffile{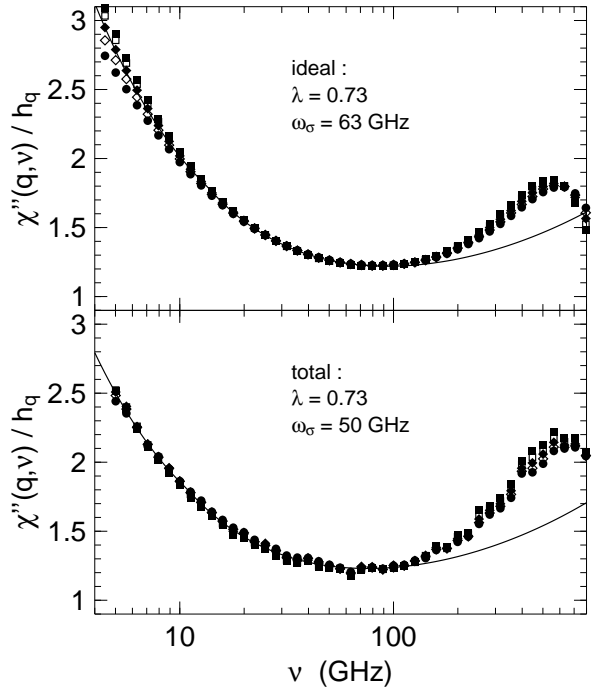}}~\\
\caption 
{Rescaled susceptibility $\chi''(q,\nu)/h_q$ as in Fig.~\protect\ref{Fbmi},
but only for the largest wavenumbers $q=1.0\ldots 1.4$~\AA$^{-1}$.
The full curves are fits with the asymptotic scaling function 
$g_\lambda(\nu/\nu_\sigma)$.
Multiple-scattering causes a shift of the minimum position
but has almost no influence on the line shape.}
\label{Ffmi}
\end{figure}

\begin{figure}
\epsfxsize=60mm\centerline{\epsffile{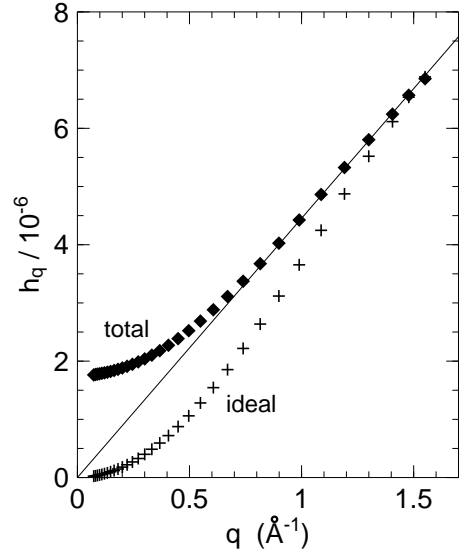}}~\\
\caption 
{Amplitudes~$h_q$ as used for the rescaling in Figs.\ \protect\ref{Fbmi}
and~\protect\ref{Ffmi}.
The wavenumber dependence is almost the same as 
for the low-frequency inelastic intensity in the {\it glass} model
[Fig.~\protect\ref{Fipi}].
The line indicates the transient linear $q$ dependence
observed in several real experiments.}
\label{Fhqm}
\end{figure}

\begin{figure}
\epsfxsize=78mm\centerline{\epsffile{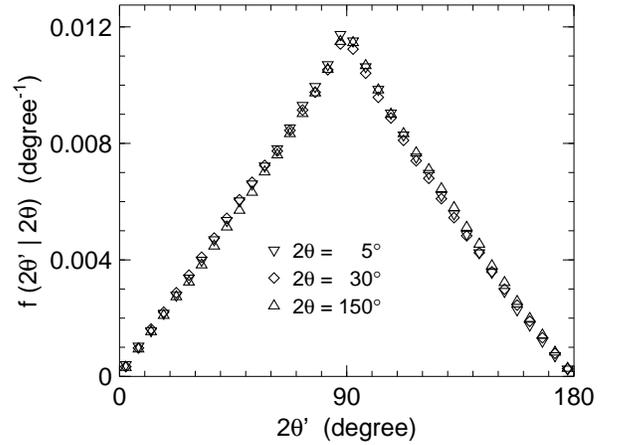}}~\\
\caption 
{Distribution $f(2\theta'|2\theta)$
of single-scattering angles $2\theta'$ 
contributing to the elastic double-scattering 
for three different detector angles $2\theta$.
From a simulation of the {\it rigid} model.
The enhanced probability of $90^\circ$ scattering events
is attributed to the sample geometry which
admits long flight paths perpendicular to the scattering plane.}
\label{Fadi}
\label{Fult} 
\end{figure}

\end{multicols}
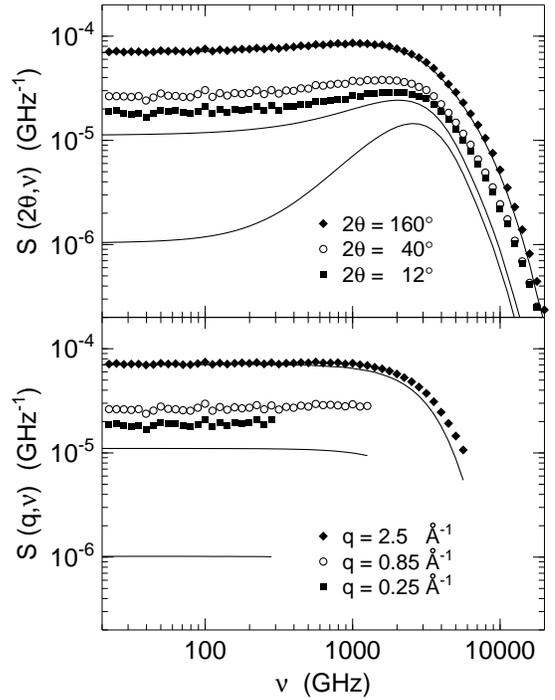
\end{document}